# Near-field Electrical Detection of Optical Plasmons and Single Plasmon Sources

Abram L. Falk[1*], Frank H. L. Koppens[1*], Chun L. Yu[2], Kibum Kang[3], Nathalie de Leon Snapp[2], Alexey V. Akimov[1], Moon-Ho Jo[3], Mikhail D. Lukin[1], and Hongkun Park[1,2]

[1]Department of Physics and [2]Department of Chemistry and Chemical Biology, Harvard University, 12 Oxford Street, Cambridge, MA 02138, USA
[3]Department of Materials Science and Engineering, Pohang University of Science and Technology, San 31, Hyoja-Dong, Nam-Gu, Pohang, Gyungbuk, 790-784, Korea
[*]These authors contribute equally to this work.

**Photonic circuits can be much faster than their electronic counterparts, but they are difficult to miniaturize below the optical wavelength scale. Nanoscale photonic circuits based on surface plasmon polaritons (SPs) are a promising solution to this problem because they can localize light below the diffraction limit[1-8]. However, there is a general tradeoff between the localization of an SP and the efficiency with which it can be detected with conventional far-field optics. Here we describe a new all-electrical SP detection technique based on the near-field coupling between guided plasmons and a nanowire field-effect transistor. We use our detectors to electrically detect the plasmon emission from an individual colloidal quantum dot coupled to an SP waveguide. The detectors are both nanoscale and highly efficient (0.1 electrons/plasmon), and a plasmonic gating effect can be used to amplify the signal even higher (up to 50 electrons/plasmon). These results enable new on-chip optical sensing applications and fulfill a key requirement for "dark" optoplasmonic nanocircuits in which SPs can be generated, manipulated, and detected without involving far-field radiation.**

SPs are charge-density waves that propagate along metal-dielectric interfaces. They can be concentrated and guided by current carrying wires, suggesting an integrated approach to optical and electrical signal processing. Our near-field plasmon detection scheme consists of an Ag nanowire (NW) crossing a Ge NW field-effect transistor (Fig. 1, Methods). The Ag NW guides[10] SPs to the Ag/Ge junction, where they are converted to electron-hole (*e-h*) pairs[11-13] and detected as current through the Ge NW. The Ag NWs are highly crystalline and defect-free[8, 14, 15], allowing SPs to propagate over distances of several microns without scattering into free-space photons. The Ge NWs are lightly *p*-doped, covered with a thin native oxide layer, and sensitive to visible light[16].

Electrical plasmon detection is demonstrated by scanning a focused laser beam across an Ag/Ge crossbar device and recording the current (*I*) through the Ge NW as a function of the diffraction-limited laser spot position. These data, recorded at $V_b = V_{gate} = 0$, show that current flows through the Ge NW only when the laser beam is focused on four distinct spots on the device (Fig. 1b). First, current is detected when the laser is focused near the Ag/Ge junction, due to the direct photoresponse of the Ge NW[16]. The photocurrent induced on the left ($I_{left}$) and right ($I_{right}$) sides of the junction have opposite signs (discussed below). Moreover, current through the Ge NW ($I_{plas}$) is recorded when the laser is focused at either end of the Ag NW.

This $I_{plas}$ signal is the key signature for electrical SP detection. Propagating SPs can be launched in the Ag NW only when the excitation laser is incident on the Ag NW ends[15]. Away from the ends, free space photon-to-SP conversion is strongly suppressed by the wave vector mismatch between the two modes. If light scattered off the Ag NW were responsible for the current flowing through the GeNW, signal would also be detected when the laser is focused on the middle of the Ag NW, in clear contrast to the data shown in Fig. 1b.

Further evidence for electrical SP detection is provided by the dependence of $I_{plas}$ on the polarization of the excitation laser. $I_{plas}$ is largest when the excitation polarization is parallel to the Ag NW axis, and smallest when perpendicular (Fig. 2b). This preference reflects the conversion efficiency of the excitation light into SP modes. The fundamental SP mode consists of cylindrically symmetric charge oscillations along the Ag NW axis[17]. The excitation of this mode is strongly suppressed when the excitation polarization is perpendicular to the wire axis. The direct Ge NW photocurrent is largest when the polarization axis is along the Ge NW (Fig. 2c), a consequence of the subwavelength NW diameter and the large dielectric contrast between the semiconductor and air[18].

An important figure of merit for our detector is the overall plasmon-to-charge conversion efficiency ($\eta$), defined as the ratio of detected charges to the number of SPs reaching the Ag/Ge junction. The values of $\eta$ in our devices typically ranged from 0.01 – 0.1 (Supplementary Information (SI) Fig. S2). This efficiency can be tuned by applying $V_{gate}$ to an additional electrical contact defined at one end of the Ag NW. In this geometry, the Ag NW is both the plasmon waveguide and a local electrical gate for the Ge nanowire (Fig. 2d, inset). As $V_{gate}$ is increased, $I_{plas}$ is enhanced (Fig. 2e), as are the magnitudes of $I_{left}$ and $I_{right}$ (Fig. 2d). Significantly, the magnitude of $I_{plas}$ and thus $\eta$ can be enhanced dramatically by applying a nonzero bias to the Ge NW, increasing 300-fold as $V_b$ is increased from 0 to 1 V (Fig. 3a). In some devices, $\eta$ exceeded 50 electrons/plasmon at $V_b = 1$ V (Fig. 3b).

These results can be understood by considering electrical plasmon detection as a multistep process. First, the AC electric field of the SP generates *e-h* pairs in the Ge NW via near-field coupling. Second, the DC electric field within the Ge NW separates these *e-h* pairs into free charges before recombination takes place. The separated *e-h* pairs are then detected as current.

The shape of the built-in DC electric potential, a potential well (Fig 1a, inset), can be inferred from the sign of the Ge photocurrent at either side of the Ag/Ge junction. The depth of this well is tuned by $V_{gate}$[16], explaining the dependence of $I_{right}$, $I_{left}$, and $I_{plas}$ on $V_{gate}$ (Figs. 2d and 2e). Asymmetric electrical contacts to the Ge can cause the potential well in the Ge to be off center with respect to the Ag NW. This asymmetry explains the difference in magnitude of $I_{right}$ and $I_{left}$, and determines the sign of the plasmon-induced current. The DC electric field in the Ge NW is nonzero even at $V_b = V_{gate} = 0$, due to charge transfer across the Ag/Ge junction[19] and/or the occupation of surface charge traps[16].

At $V_b = 0$, $\eta$ is the genuine SP detection efficiency. The amplification of this plasmon detection signal at nonzero $V_b$ is due to a plasmon-induced gating effect, similar to the photo-gating effect observed previously in Ge NWs[16]. When incident light or SPs excite *e-h* pairs in the *p*-doped Ge NW, a portion of the minority carriers (electrons) migrate to the NW surface and get trapped. These trapped charges gate the semiconductor and increase the conductance, thereby producing electrical gain. The sublinear power dependence—$\eta$ decreases as the excitation laser power increases (Fig. 3b)—reflects the saturation of surface charge traps on the Ge NW.

Finite-difference time-domain (FDTD) simulations of our devices provide further insight into the device operation. In these simulations, an SP pulse is launched in the fundamental mode of the Ag NW. As it propagates, the electric field intensity (Fig. 3c) and Poynting vector (Fig. 3d) are monitored as functions of position and time. Once the SP flux reaches the Ge/Ag junction, it can be reflected, transmitted, absorbed by the Ge NW, or scattered to the far field. The simulation results show that a 40-nm diameter Ge NW absorbs 20 (50)% of the SP flux when the Ag NW diameter is 100 (50) nm (Fig. 3d and SI Fig. S3). This large absorption fraction originates from the high absorption constant[19] of Ge and the strong SP confinement, which increases for smaller Ag NW diameters.

A comparison of the simulated Ge absorption fraction (20%) and the experimentally measured $\eta$ values (1-10% at $V_b = 0$) suggests that SP absorption by the Ge NW is not the main factor that limits our detectors' efficiencies. We note that an electric field profile that is perfectly symmetric with respect to the Ag-NW axis would not result in a net current flow at $V_b = 0$ when cylindrically symmetric SPs impinge on the Ag/Ge junction. As such, the sign and magnitude of $I_{plas}$ reflect the contact asymmetry at the Ag/Ge junction, and properly designed asymmetric gates could enhance the intrinsic efficiency.

We demonstrate the utility of our near-field SP detector by electrically detecting emission from a CdSe quantum dot acting as a single plasmon source (Fig. 4). The tight field confinement around Ag NWs causes a large fraction of the spontaneous emission from nearby emitters (i.e. 30-100 nm away) to be captured as SP modes[9, 20, 21]. These SPs are then converted into an electrical signal at a Ge NW detector. Individual quantum dots (QDs) are coupled to a Ag NW by covering an Ag/Ge device with a 30 nm film of poly(methyl methacrylate) containing a dilute concentration of chemically synthesized CdSe QDs. Optical fluorescence measurements (Fig. 4a) show that some QDs are close to the Ag NW. When the laser excites one of these QDs, a current signal in the Ge NW detector is observed ($I_1$, circled in Fig. 4b) in addition to optical fluorescence from the QD.

The dependence of this signal on the excitation wavelength ($\lambda_{ex}$) is a clear proof that the current signal results from QD emission into SPs (Fig. 4c). This spectrum, a downward trend with increasing $\lambda_{ex}$ and a distinct peak at 655 nm, closely resembles the absorption spectrum of CdSe QDs[22]. In contrast, when plasmons are launched from a NW end by direct photon-to-SP conversion ($I_2$, circled in Fig. 4b), the spectrum displays a gradual upward trend with increasing $\lambda_{ex}$, as expected from the longer SP propagation lengths at higher wavelengths[1]. Significantly, photon correlation measurements of the far-field fluorescence (Fig. 4d) demonstrate a clear anti-bunching signature, indicating that this spot corresponds to an individual QD. Combined with the fact that no signal is recorded when the laser is focused on QDs far from the NW, these measurements prove that single QD emission captured into SPs can be detected electrically without intermediate far-field photons (See SI S4 for a discussion of the sign of $I_1$ and $I_2$ and an additional QD detection data set).

When the QDs are excited at $\lambda_{ex} = 500$ nm, we find $I_{plas} \sim 1$ pA at $V_b = 0$, corresponding to ~6 × $10^6$ electrons/sec. Assuming that the QD emission directed into plasmons equals its far-field emission, a typical QD can generate up to $5 \times 10^7$ plasmons/sec[9]. This simple consideration implies that the detection efficiency of the QD emission is ~0.1 electrons/plasmon, consistent with the $\eta$ value of this device at $V_b = 0$ (SI Fig. S2).

Nanoscale near-field SP detection opens up several directions for further research. In conjunction with an electrically driven plasmon source[23], a near-field SP detector could be integrated into a "dark" optoelectronic-plasmonic nanocircuit in which all coupling is in the near field. Moreover, the plasmon detection sensitivity could be improved by using a nanoscale avalanche photodiode[24] as the SP detector, potentially enabling efficient electrical detection of individual plasmons. Electrical plasmon detectors could lead to new applications for optical sensing without collection optics, including the measurement of plasmon states whose coupling to the far field is suppressed by symmetry[25]. Finally, the strong near-field coupling between single-plasmon emitters and plasmonic nanocircuits could lead to completely new capabilities that are not available with conventional photonics, such as nonlinear switches, single-photon transistors, and quantum non-demolition detectors[20, 26, 27].

**Figures**

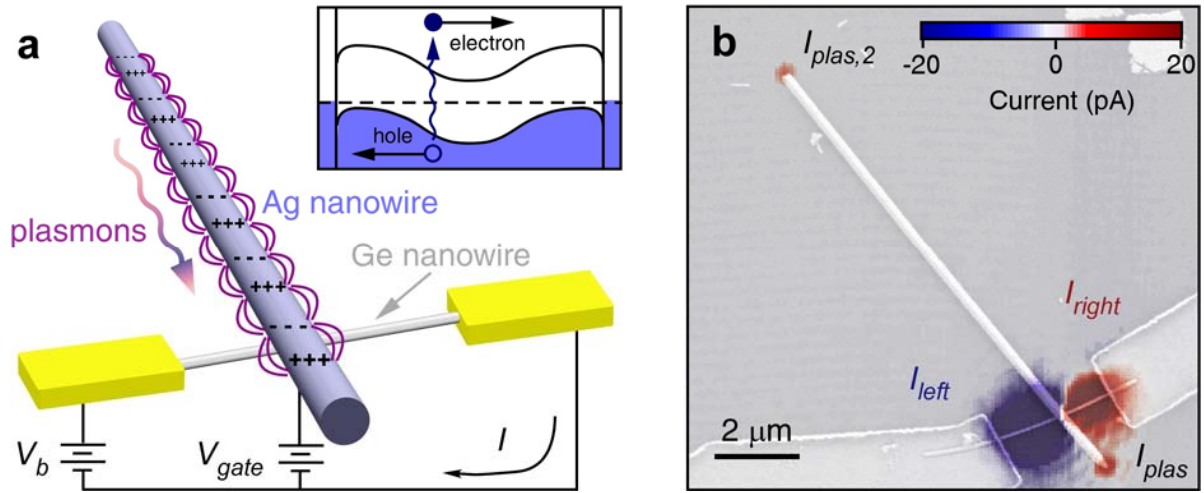

**Figure 1 | Electrical Plasmon Detection. a,** Schematic of electrical plasmon detector operation. Inset: electron-hole pair generation and separation in the Ge NW detector. **b,** Scanning electron micrograph of Device 1, overlaid with the current through the Ge NW as a function of excitation laser position. Excitation laser power $P$ = 2.0 μW, wavelength $\lambda_{ex}$ = 532 nm, $V_b$ = 0.

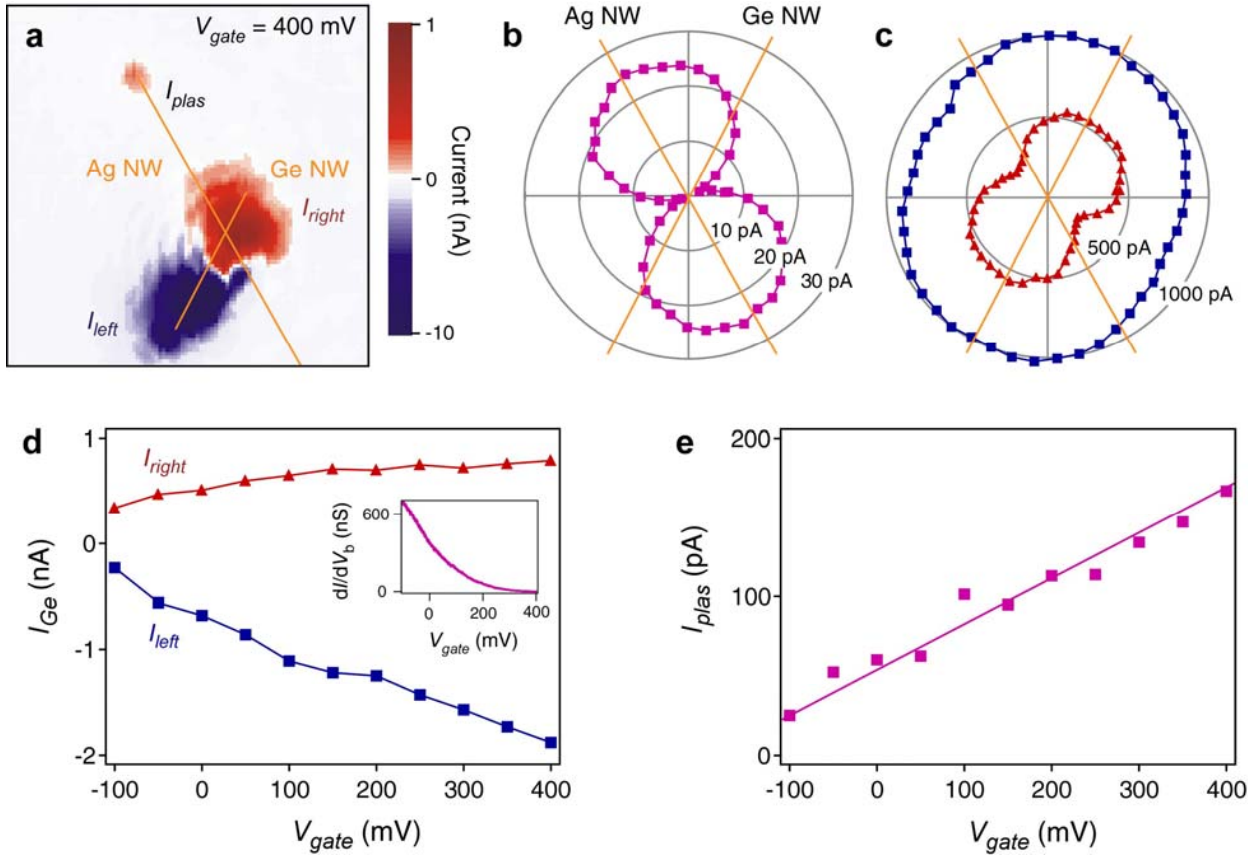

**Figure 2 | Polarization and Gate Effects on Plasmon Detection. a,** $I$ as a function of laser spot position for device 2. $V_{gate}$ = 400 mV, $V_b$ = 0, $P$ = 3.4 μW, $\lambda_{ex}$ = 600 nm. **b,** $I_{plas}$ and **c,** $I_{right}$ (red triangles) and $I_{left}$ (blue squares) as a function of excitation light polarization. **d,** $I_{right}$ (red triangles) and $I_{left}$ (blue squares) as a function of $V_{gate}$. Inset: Gate response of the Ge NW conductance in nanosiemens. **e,** $I_{plas}$ as a function of $V_{gate}$.

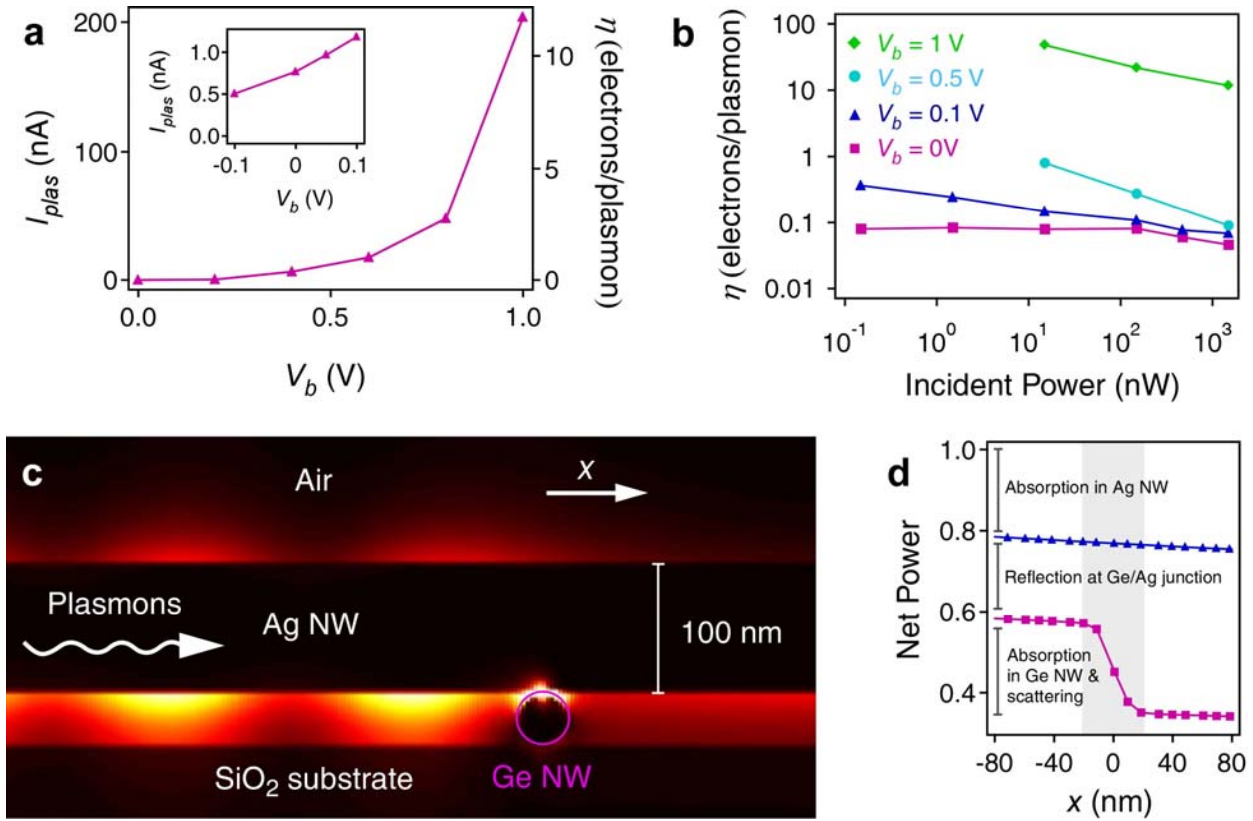

**Figure 3 | Gain in Plasmon Detectors and Simulation. a,** $I_{plas}$ as a function of $V_b$ for Device 3. $P = 1.5$ µW, $V_{gate} = 0$, $\lambda_{ex} = 600$ nm, Ag NW diameter = 100 nm, Ge NW diameter = 40 nm. Inset: $I_{plas}$ as a function of $V_b$ for smaller values of $V_b$. **b,** $\eta$ as a function of $P$ and $V_b$. **c,** Finite element time domain simulation of electric field intensity. Ag NW diameter = 100 nm, Ge NW diameter = 40 nm. **d,** Net power in the plasmon mode as a function of distance ($x$) along the Ag NW from the Ge NW center. At $x = 0$, the net power drops in the Ag/Ge device (violet squares) due to absorption in the Ge NW and scattering, whereas there is no drop in a simple Ag NW with no Ge crossbar (blue triangles). The shaded area represents the position of the Ge NW. Approximately 30% of the SP energy is reflected, 3% scatters to the far field, 20% is absorbed by the Ge NW, and the rest is transmitted.

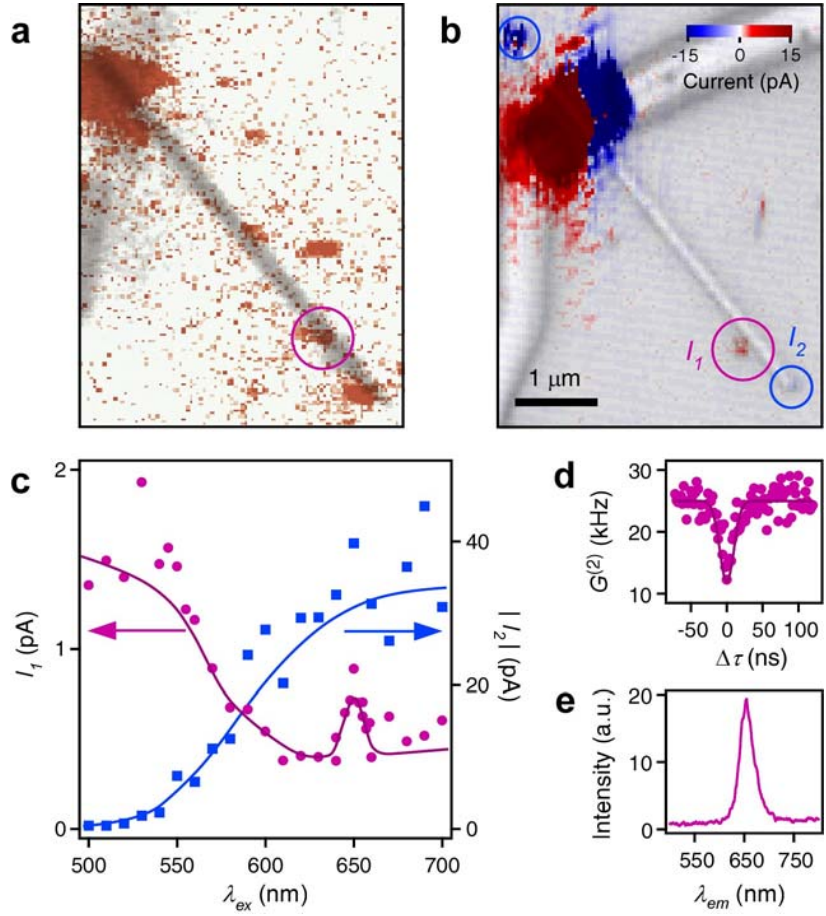

**Fig. 4 | Electrical detection of emission from a single CdSe colloidal QD. a,** Optical emission from QDs, measured with a confocal microscope, overlayed on a reflection image. The emission is filtered with a 600 nm long-pass filter. **b,** $I$ as a function of laser position, overlaying a reflection image of Device 4. $P = 1.5$ µW, $\lambda_{ex} = 530$ nm. $V_b = 0$, $V_{gate} = 0$. The violet circle indicates the detector signal corresponding to QD emission into the Ag NW plasmon modes. The blue circles correspond to detection of surface plasmons launched directly by the laser, at the ends of the NW. **c,** $I_1$ (QD emission, violet circles) and $I_2$ (direct laser-to-plasmon signal, blue squares) as a function of $\lambda_{ex}$. **d,** Second-order self-correlation function $G^{(2)}(\tau)$ of the QD fluorescence. The number of coincidences at $\tau = 0$ approaches the 0.5 mark, confirming that a single QD is present. **e,** QD emission spectrum.

**Methods**

The Ag nanowires (NWs) used in our experiments were synthesized using a modified polyol method described previously[14]. The Ge NWs were synthesized using a vapor-liquid-solid chemical vapor deposition technique[28-30]. They were catalyzed by Au nanoparticles, grown in $GeH_4/H_2$ gas, and subsequently doped to be *p*-type in $B_2H_4/H_2$ gas.

The SP detectors were fabricated by spin casting Ge NWs and Ag NWs on a degenerately doped Si substrate with 300 nm $SiO_2$ grown on top. Once the Ag/Ge NW crossbars were identified, the Ge NWs were electrically contacted using conventional electron beam lithography and Ti/Au metal deposition (15 nm/150 nm).

The excitation of SPs was performed using a Koheras SuperK supercontinuum laser. The laser was coupled to an acousto-optic tunable filter, enabling the excitation wavelength to be selected. The beam was spatially filtered through a pinhole, and then directed to an optical microscope using a scanning mirror. The objective lens of the microscope (100x, 0.8 numerical aperture) focused the beam to a diffraction-limited spot on the device. The details of the photon correlation measurement used for measuring the quantum dot emission statistics are given elsewhere[9].

**Supplementary Information** accompanies this paper.

**Acknowledgements** We would like to acknowledge Darrick Chang for valuable discussions and Murray McCutcheon for assistance with the FDTD simulations. This work was supported by the Defense Advanced Research Projects Agency, the National Science Foundation, the Air Force Office of Scientific Research, and Samsung Electronics.

**Author Information** The authors declare no competing financial interests. Correspondence and requests for materials should be addressed to H.P. (Hongkun_Park@harvard.edu) and M.D.L. (lukin@fas.harvard.edu).

**Supplementary Information for Near-field Electrical Detection of Optical Plasmons and Single Plasmon Sources**

Abram L. Falk[1*], Frank H. L. Koppens[1*], Chun L. Yu[2], Kibum Kang[3], Nathalie de Leon Snapp[2], Alexey V. Akimov[1], Moon-Ho Jo[3], Mikhail D. Lukin[1], and Hongkun Park[1,2]

[1]Department of Physics and [2]Department of Chemistry and Chemical Biology, Harvard University, 12 Oxford Street, Cambridge, MA 02420, USA

[3]Department of Materials Science and Engineering, Pohang University of Science and Technology, San 31, Hyoja-Dong, Nam-Gu, Pohang, Gyungbuk, 790-784, Korea

[*]These authors contribute equally to this work.

**S1 Propagation Length and Input Coupling of Silver NW SP Modes**

Simulated and experimentally determined values of SP propagation length ($L_0$) are shown in Fig. S1a. The Ag NWs with a smaller diameter ($d_{Ag}$) exhibit smaller $L_0$ values due to the tighter field confinement of their SP modes[1]. The propagation length simulations were performed with the Ag NW resting on an $SiO_2$ substrate, so that the simulated $L_0$ values include substrate ($SiO_2$) effects. The experimental $L_0$ values are consistently smaller than the calculated values due to imperfections in the Ag NWs, as observed previously[2].

The experimental $L_0$ values were estimated in two ways. The filled squares in Fig. S1a are from electrical plasmon detection measurements in which both ends of the Ag NW were accessible (Fig. S1b). Assuming that photon-to-SP conversion is identical at the two NW ends, the difference in $I_{plas}$ at those two points reflects the difference in the propagation loss, allowing $L_0$ to be determined. The open square in Fig. S1a is from an exponential fit of the emission intensities from a series of defects in a given NW, following Sanders *et al.*[2]. A linear fit to these data gives $L_0 = 26.7 d_{Ag}$. This fit applies in the window 80 nm < $d_{Ag}$ < 250 nm.

The probability that a photon from the objective is scattered into a SP was estimated by focusing the excitation laser at one end of an Ag NW, and then detecting the far-field emission caused by SP scattering at the opposite end of the NW using a charge-coupled device (CCD) camera (Fig. S1c). The magnitude of this emission depends both on the photon-to-SP conversion efficiency ($\alpha_{in}$) and the SP-to-photon conversion efficiency ($\alpha_{out}$) at the Ag NW ends. Taking the SP propagation loss into account, the measured photon flux emitted from an Ag NW ($P_{out}$) is given by:

$$P_{out} = \alpha_{out} e^{-L/L_0} \alpha_{in} P \qquad \text{(S1)}$$

Here, *P* is the incident laser flux (measured underneath the microscope objective), and *L* is the length of the Ag NW. Since our laser spot is diffraction limited, and thus single mode, it is reasonable to approximate that $\alpha_{in} = \alpha_{out}$. Then, fitting data from multiple Ag NWs to Eq. S1, the value of $\alpha_{in}$ was determined to be 0.05 ± 0.016. This value of $\alpha_{in}$ was determined from the measurements of 30 NWs that exhibited good SP propagation.

Interestingly, the value of $\alpha_{in}$ was found not to depend sensitively on $d_{Ag}$ within the range 70 < $d_{Ag}$ < 250 nm. In principle, higher $\alpha_{in}$ values are expected for larger diameter Ag NWs, since the SP modes of those wires have a smaller *k*-vector mismatch to free space. The most likely reason for the lack of strong dependence is that actual Ag NWs are often tapered at the ends (Fig. S1d).

**S2 Calculation of the SP-to-Charge Conversion Efficiency**

The SP-to-charge conversion ratio ($\eta$) is defined to be the ratio of detected charges to SPs that reach the Ge/Ag junction. This fraction can be calculated from the incisdent power using:

$$\eta = \left(\frac{I_{plas}}{e}\right)\left(\frac{hc/\lambda_{ex}}{e^{-x/L_0}\alpha_{in} P}\right), \qquad \text{(S2)}$$

where $e$ is the electron charge, $hc/\lambda_{ex}$ is the excitation photon energy, and $x$ is the Ag wire length from the excitation end to the Ge detector.

The values of $\eta$ for several devices are shown in Fig. S2, calculated using Eq. S2, $L_0 = 26.7d_{Ag}$, and $\alpha_{in} = 0.05$. These measurements were performed at $V_b = 0$, thus representing a true SP detection efficiency of the device. As mentioned in the main text, $\eta$ was found to decrease as $P$ is increased, due to photo/SP-generated charges screening the DC electric field in the Ge.

**S3 Simulation of Plasmon Detection**

Finite-difference time-domain (FDTD) simulations of Maxwell's equations were performed using the software package Lumerical (Fig. S3 and Fig 3c-d of the main text). The simulated device is a 3 μm long Ag NW ($\varepsilon = -16.1+0.05i$ when $\lambda_{ex} = 600$nm[3]) lying on top of a 1 μm long Ge NW ($\varepsilon = 4.6+1.9i$ for $\lambda_{ex} = 600$nm[3]), which in turn is resting on a SiO$_2$ substrate (Fig S3a). The NW diameters match those in Device 3: $d_{Ag} = 100$ nm, $d_{Ge} = 40$ nm. A mode source excites the fundamental SP mode of the Ag NW. The excitation pulse is 4 fs long and has a center frequency of 500 THz, corresponding to $\lambda_{ex} = 600$ nm in vacuum. The propagation of this mode along the Ag NW and its interaction with the Ge NW is simulated for a 30 fs interval. This interval is long enough to include reflections at the Ge/Ag crossing point and absorption in Ge NW but short enough to exclude the effects of reflections at the Ag NW end.

The analysis of the simulation results was performed in the frequency domain. Figure 3c of the main text shows the electric field intensity of a cross-section cut through the Ag NW axis. The standing wave pattern on the anterior (left) side of the Ge/Ag crossing point is due to reflections at the crossing point. A strong reduction of the electric field intensity inside the Ge NW is observed. This is due both to depolarization and absorption within the Ge, corresponding to the real and imaginary components of the Ge index of refraction. A considerable fraction of the SP flux is absorbed.

In order to extract the energy fraction that the Ge NW absorbs, the Poynting vector of the SP was monitored at different positions along the Ag wire, including just before and after the Ge/Ag crossing point (Fig 3d in the main text). The difference between these two constitutes three different loss processes: (1) absorption in the Ag (SP propagation loss), (2) far-field scattering due to the rapid change in the dielectric environment, and (3) absorption by the Ge. The SP propagation loss is known from simulations within the Ag NW. The far-field scattering is extracted by monitoring the power propagating perpendicular to the Ag NW (Fig. S3b). The Ge NW absorption is the remainder. When $d_{Ag}$=100 nm and $d_{Ge}$ = 40 nm, 30% of the SP energy is reflected, 3% scatters to the far field, 20% is absorbed by the Ge NW, and the rest is transmitted (Fig. 3d in main text). Absorption increases as $d_{Ag}$ decreases (Fig. S3b), due to increased SP field confinement.

Because this simulation models only the absorption of SPs by the Ge NW and not the separation or diffusion of *e-h* pairs, the simulated absorption fraction should be considered as an upper limit for the measured $\eta$ at $V_b$ = 0. We note that an electric field profile that is perfectly symmetric with respect to the Ag-NW axis would not result in a net current flow at $V_b = 0$ when cylindrically symmetric SPs impinge on the Ag/Ge junction. As such, the sign and magnitude of $I_{plas}$ reflect the contact asymmetry at the Ag/Ge junction. Properly designed asymmetric gates could potentially enhance the intrinsic efficiency.

**S4 Additional Electrical Detection of Quantum Dot Data and Analysis**

Figure S4 shows an additional device in which near-field detection of quantum dot emission is observed. Unlike device 4, presented in Fig. 4 of the main text, we could observe no anti-bunching from the electrically detected emitter (circled in Fig. S4b). This observation indicates that the emitter is a quantum dot cluster, not a single quantum dot. In this device, the location of the quantum cluster coincides with a nanoparticle in the SEM image. This nanoparticle is quite likely the quantum dot cluster itself. The photocurrent at this spot is due to quantum dot emission, as confirmed by the excitation spectrum recorded by the Ge plasmon detector (Fig S4d).

In Fig. 4b of the main text, the sign of $I_1$ differs from that of $I_2$. The reasons for this are not fully understood. One possibility is that photogenerated carriers charge the Ag NW and/or the Ge-NW surface traps as the excitation laser is scanned over the device and change the electric field profile in the Ge NW, a phenomenon that was observed in some devices. Another possibility is that the QD emission couples into a different plasmon mode, which couples differently to the Ge NW.

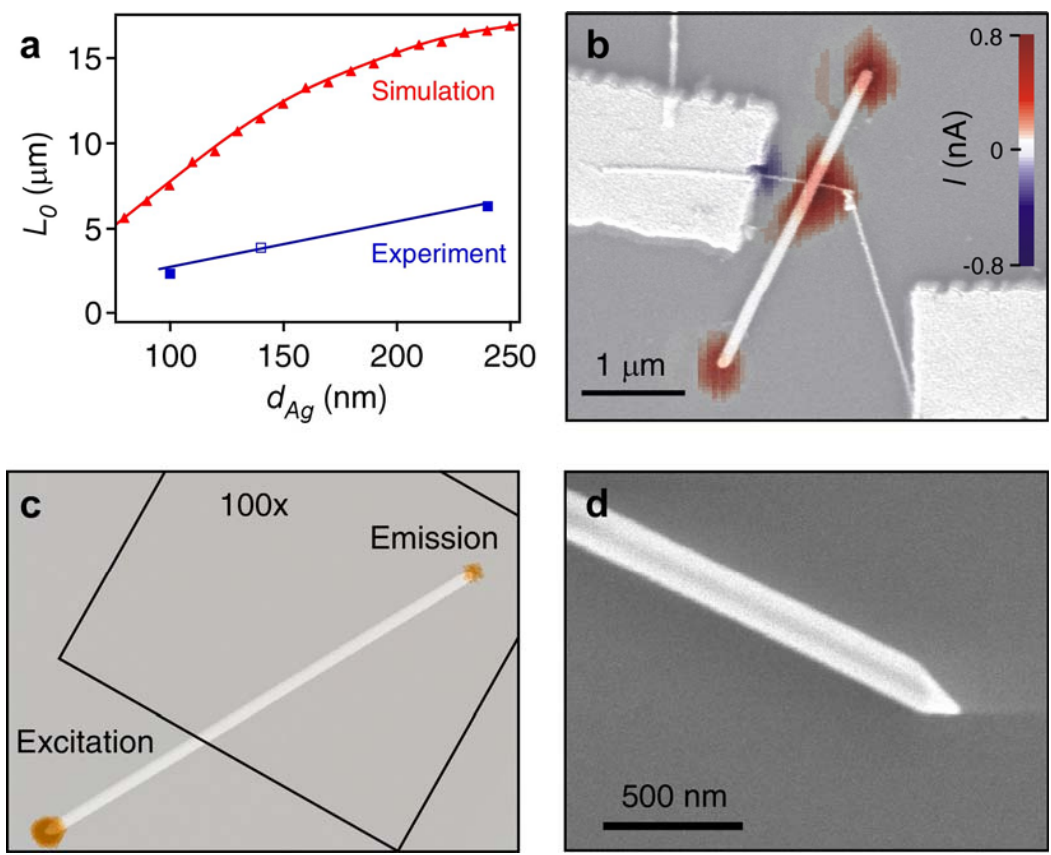

**Fig. S1 | Propagation Length and Coupling Factor Measurements. a,** $L_0$ as a function of $d_{Ag}$. The red triangles are from FDTD simulations. The filled blue squares are inferred $L_0$ values from electrical detection data. The open blue square is from fitting an exponential to the emission from an Ag NW with several small defects. **b,** A photocurrent image of Device 3 superimposed over an SEM image. $V_b = V_{gate} = 0$, $\lambda_{ex} = 600$ nm. **c,** A CCD image of an Ag NW being excited at one end and emitting light at the other, superimposed over an SEM image of the same NW. The intensity scale is 100x in the rectangle surrounding the emission end. **d,** An SEM image of an Ag NW, showing a tapered tip.

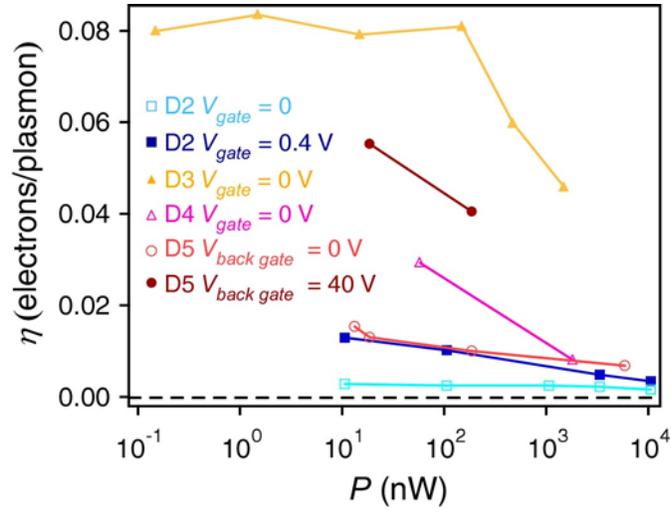

**Fig. S2 | SP-to-charge conversion efficiency ($\eta$)** for several of the devices discussed in the main text at $V_b = 0$, $\lambda_{ex} = 600$ nm.

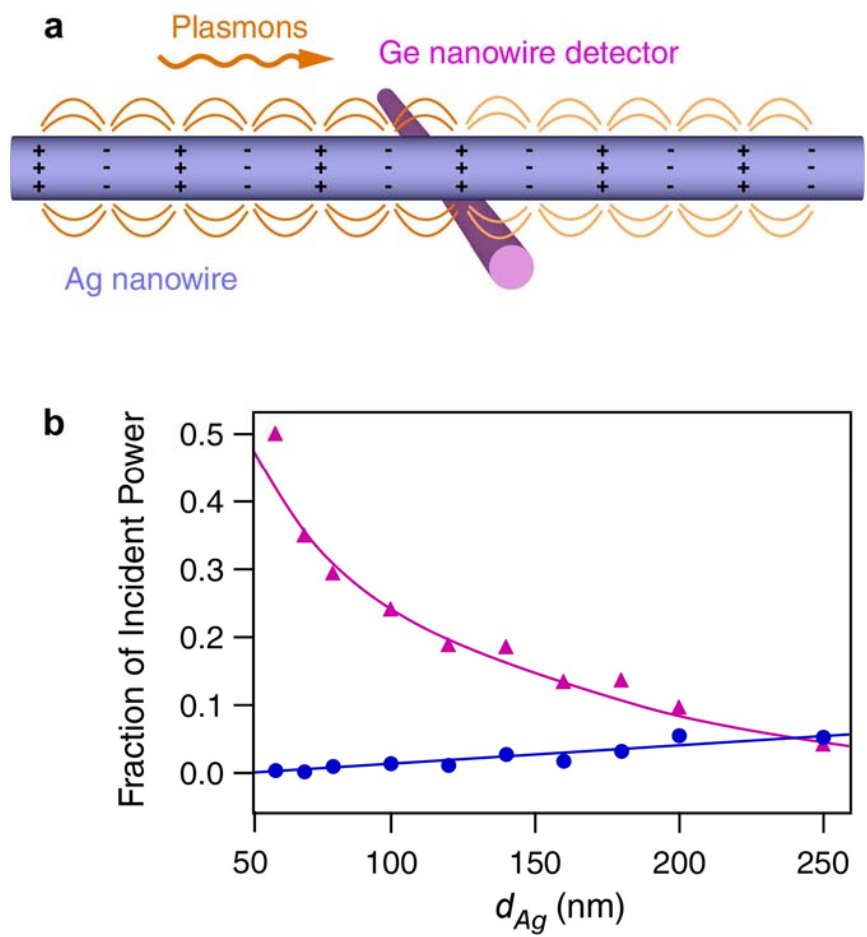

**Fig. S3 | FDTD Simulation. a,** A diagram of the NW geometry that was simulated. The Ag and Ge NWs are 3 μm and 1 μm long, respectively. SPs are launched in the fundamental mode of the Ag NW and get partially absorbed by the Ge NW. The NW diameters were chosen to match Device 3: $d_{Ag}$ = 100 nm, $d_{Ge}$ = 40 nm. **b,** SP absorption (violet triangles) and scattering to far field (blue circles) as a function of $d_{Ag}$. $d_{Ge}$ = 40 nm. See also Fig. 3 in the main text.

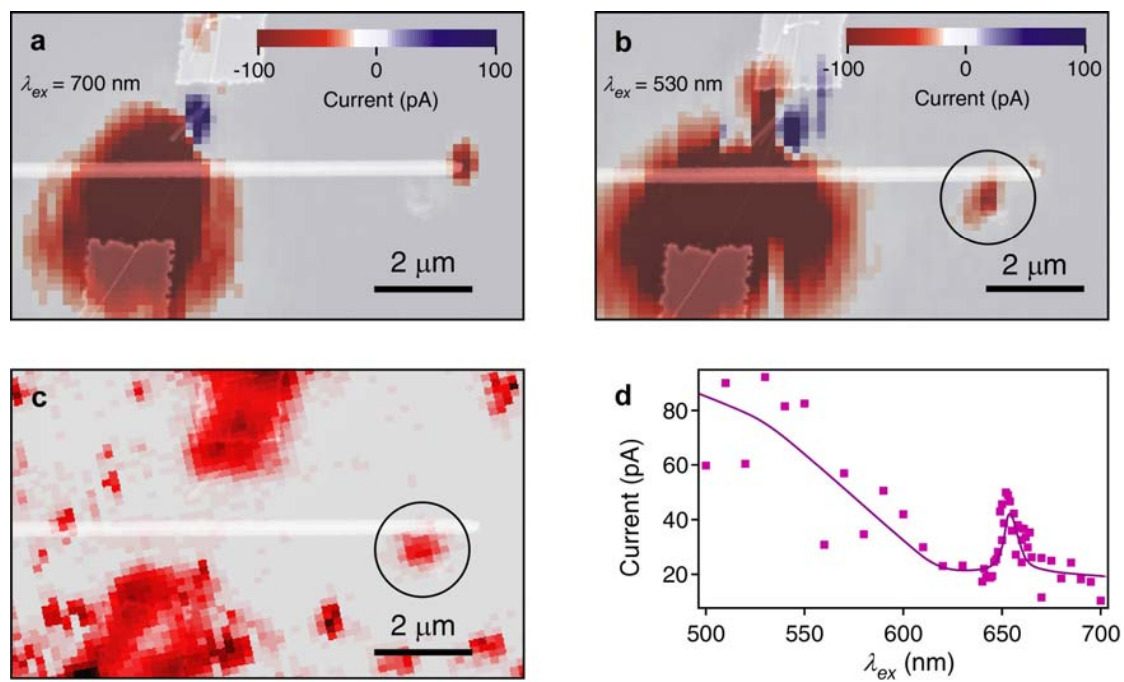

**Fig. S4 | Electrical detection of a quantum dot cluster. a,** Photocurrent image of Device 6 at $\lambda_{ex}$ = 700 nm, $V_b$ = 0, overlaying an SEM image of the device. **b,** Photocurrent image of the same device at $\lambda_{ex}$ = 532 nm, $V_b$ = 0, overlaying an SEM image of the device. **c,** Optical emission image, $\lambda_{ex}$ = 532 nm, filtered with a longpass filter at $\lambda$ = 600 nm. **d,** Excitation spectrum of the SP electrical detection corresponding to the circled point Figs. S4b and S4c. The line is a guide to the eye.